\documentclass{emulateapj}

\pdfoutput=1

\usepackage{amsmath}
\usepackage{amssymb}
\usepackage{epsfig}
\usepackage{graphicx}
\usepackage{color}

\def\gtaprx {\lower .1ex\hbox{\rlap{\raise .6ex\hbox{\hskip .3ex
	{\ifmmode{\scriptscriptstyle >}\else
		{$\scriptscriptstyle >$}\fi}}}
	\kern -.4ex{\ifmmode{\scriptscriptstyle \sim}\else
		{$\scriptscriptstyle\sim$}\fi}}}
\def\ltaprx {\lower .1ex\hbox{\rlap{\raise .6ex\hbox{\hskip .3ex
	{\ifmmode{\scriptscriptstyle <}\else
		{$\scriptscriptstyle <$}\fi}}}
	\kern -.4ex{\ifmmode{\scriptscriptstyle \sim}\else
		{$\scriptscriptstyle\sim$}\fi}}}

\newcommand{\cutt}[1]{\textcolor{blue}{}}

\newcommand{\Ms}{{\ensuremath{M_{\odot} }}}

\begin{document}

\title{Detecting Ancient Supernovae at $z \sim$ 5 - 12 with {\it CLASH}}

\author{Daniel J. Whalen\altaffilmark{1}, Joseph Smidt\altaffilmark{2}, Claes-Erik 
Rydberg\altaffilmark{3}, Jarrett L. Johnson\altaffilmark{1}, Daniel E. 
Holz\altaffilmark{4} and Massimo Stiavelli\altaffilmark{5}}

\altaffiltext{1}{Institute of Cosmology and Gravitation, Portsmouth University, Dennis 
Sciama Building, Portsmouth PO1 3FX}

\altaffiltext{2}{Los Alamos National Laboratory, Los Alamos, NM 87545}

\altaffiltext{3}{Universit\"{a}t Heidelberg, Zentrum f\"{u}r Astronomie, Institut f\"{u}r 
Theoretische Astrophysik, Albert-Ueberle-Str. 2, 69120 Heidelberg, Germany}

\altaffiltext{4}{Enrico Fermi Institute, Department of Physics, and Kavli Institute for 
Cosmological Physics, University of Chicago, Chicago, IL 60637, USA}

\altaffiltext{5}{Space Telescope Science Institute, 3700 San Martin Drive, Baltimore, 
MD 21218}

\altaffiltext{6}{CCS-2, Los Alamos National Laboratory, Los Alamos, NM 87545}

\begin{abstract}

Supernovae are important probes of the properties of stars at high redshifts because 
they can be detected at early epochs and their masses can be inferred from their light 
curves. Finding the first cosmic explosions in the universe will only be possible with the 
{\it James Webb Space Telescope}, the Wide-Field Infrared Survey Telescope and the 
next generation of extremely large telescopes.  But strong gravitational lensing by 
massive clusters, like those in the {\it Cluster Lensing and Supernova Survey with 
Hubble} ({\it CLASH}), could reveal such events now by magnifying their flux by factors 
of 10 or more.  We find that {\it CLASH} will likely discover at least 2 - 3 core-collapse 
supernovae at 5 $< z < $ 12 and perhaps as many as ten.  Future surveys of cluster 
lenses similar in scope to \textit{CLASH} by the {\it James Webb Space Telescope} 
might find hundreds of these events at $z \lesssim$ 15 - 17.  Besides revealing the 
masses of early stars, these ancient supernovae will also constrain cosmic star 
formation rates in the era of first galaxy formation.

\vspace{0.1in}

\end{abstract}


\section{Introduction}

In recent years, Type Ia supernovae (SNe) at $z > $ 0.1 have been the focus of 
much attention, and rightly so for their potential to trace cosmic acceleration and 
constrain the dark energy equation of state.  But with the advent of the {\it James 
Webb Space Telescope} ({\it JWST}) and the next generation of extremely large 
telescopes, it will soon be possible to detect SNe at the edge of the observable 
universe at $z \sim$ 10 - 20 and use them to probe the earliest stellar populations.  
These stars are key to the properties of primeval galaxies \citep{fg11}, the origins 
of supermassive black holes \citep[e.g.,][]{jlj12a,jet13,latif13c,latif13a,schl13} and 
the reionization and chemical enrichment of the early intergalactic medium 
\citep[IGM;][]{wan04,bc05,fet05,jet09b,ritt12,ss13} \citep[for recent reviews, see][]{
dw12,glov12}.

Supernovae can reveal the properties of stars at this era because they are visible 
at high redshifts and their masses can be inferred from their light curves.  Recent 
calculations show that pair-instability (PI) and pulsational pair-instability (PPI) SNe 
are visible in the near infrared (NIR) at $z \sim$ 20 to {\it JWST} and the Thirty
Meter Telescope (TMT) \citep{jw11,kasen11,wet12a,hum12,wet13d}.  These 
explosions will also be visible at $z \sim$ 15 - 20 to the Wide-Field Infrared Survey 
Telescope (WFIRST) and the Wide-Field Imaging Surveyor for High Redshift 
(WISH), which could harvest these events in large numbers because of their wide 
search fields.  It is now known that core-collapse (CC) SNe will be detected at $z 
\sim$ 10 - 15 by {\it JWST}, at $z \sim$ 7 - 10 by WFIRST and WISH, and at $z <$ 
7 by {\it Euclid} \citep{wet12c,mw12}.  Supermassive SNe \citep{wet12d,jet13a,
wet13a,wet13b} and Type IIn SNe \citep{moriya12,wet12e,tet12} will also be visible 
to these missions at $z \gtrsim$ 15. 

But could such events be detected in surveys now?  In principle, strong lensing 
by massive galaxy clusters at $z \lesssim 1$ could boost flux from SNe by factors 
of 10 or more and allow them to be detected at $z \gtrsim$ 10 by current 
instruments.  \citet{om10} have estimated detection rates for strongly lensed SNe 
at low redshifts in future all-sky surveys by the Large Synoptic Survey Telescope 
(LSST), and \citet{aman11} have discovered a lensed SN at $z \sim$ 1.7 at the 
edge of the massive galaxy cluster A1689.  Many such clusters have now been 
targeted by the Cluster Lensing and Supernova Survey with {\it Hubble} ({\it 
CLASH}) to find Type Ia SNe at $z \sim$ 2 \citep{clash}. \citet{pa13} have already 
found three SNe at $z < $ 0.4 in {\it CLASH}. The {\it Hubble Space Telescope} 
({\it HST}) Frontier Fields program will soon examine six additional cluster lenses 
to hunt for early galaxies.      

We have now calculated the number of SNe that might be discovered at $5 < z < 
20$ in {\it CLASH} and by future cluster surveys by {\it JWST}.  We describe our 
lens model and derive lensing volumes as a function of redshift in Section 2.  In 
Section 3, we provide NIR light curves for 15 - 40 \Ms\ CC SNe.  Star formation 
rates at 5 $ < z < $ 25 compiled from gamma-ray burst (GRB) rates, high-redshift 
galaxy luminosity functions, and numerical simulations are presented in Section 4.  
In Section 5, these rates are convolved with lensing volumes and SN luminosities 
to determine the rates at which SNe will be found by a single cluster lens as a 
function of redshift.  We conclude in Section 6.
  
\section{Lensing Model} 

\begin{figure}
\plotone{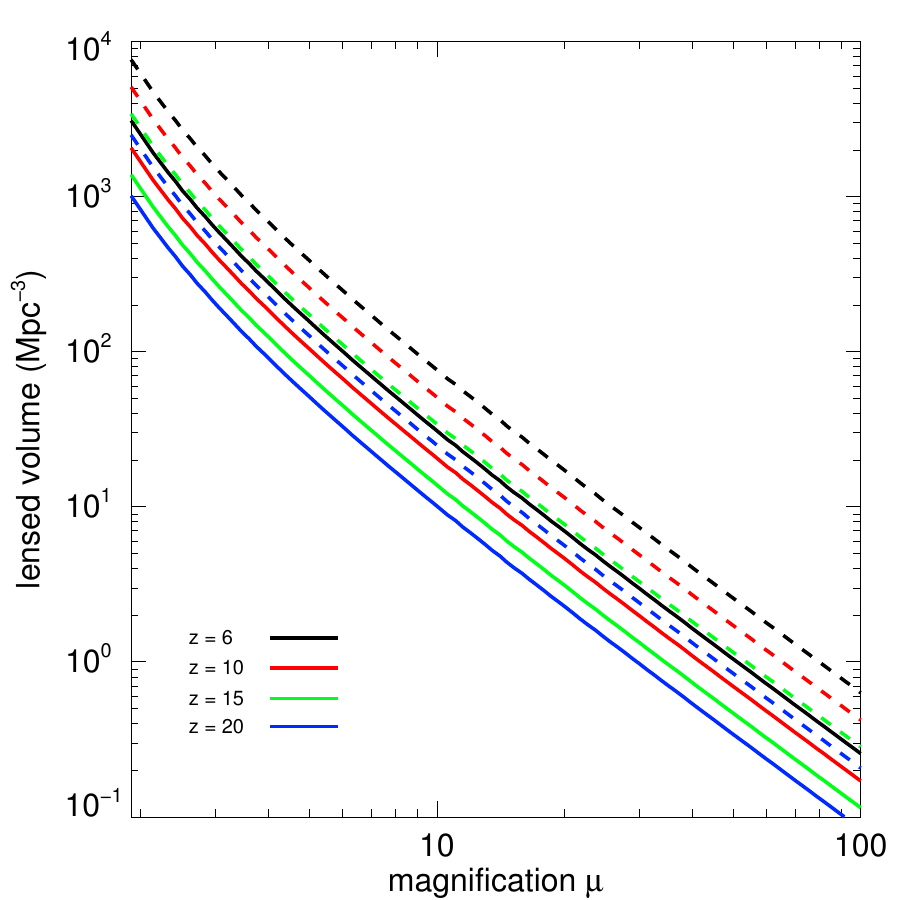}
\caption{Volumes lensed by magnifications greater than $\mu$, plotted as a 
function of $\mu$ for 35 arcsec (solid) and 55 arcsec (dashed) cluster lenses.} 
\vspace{0.1in}
\label{fig:mu}
\end{figure}

In lieu of actual lensing maps for each cluster \citep{zit09}, which are not yet 
available for {\it CLASH}, we model a cluster lens as a singular isothermal sphere 
whose magnification, $\mu$, is uniquely specified by its Einstein radius, $\theta_
\mathrm{E}$.  We use equations 4 - 6 from \citet{pl13} to compute the area in the 
source plane that is magnified above a given $\mu$ with cosmological parameters 
from the most recent Planck$+$WP$+$highL$+$BAO data:  $\Omega_\mathrm{M} 
= $ 0.308 and $\Omega_\Lambda = $ 0.692 \citep{planck}.  Lensing volumes as a 
function of $\mu$ and $z$ for $\theta_\mathrm{E} =$ 35 and 55 arcsec are plotted 
in Fig.~\ref{fig:mu}.  These two radii bracket $\theta_\mathrm{E}$ for 5 of the 25 
clusters selected for \textit{CLASH} and are typical of clusters with high $\mu$ (the 
other 20 cluster lenses in the survey have $\theta_{\mathrm{E}} \sim$ 10 - 30 
arcsec).  Note that lensing volumes decrease with redshift above $z \sim$ 2 
because the luminosity distance $D_{\mathrm{A}}$ peaks and begins to fall above 
this redshift.  This trend, plus the fact that star formation rates plummet at high $z$
as we show later, are the two greatest limitations to using cluster lenses to detect
early SNe.    

Real clusters are better described by Navarro-Frenk-White (NFW) profiles.  To 
check the validity of the SIS approximation, we have redone the analysis of \citet{
pl13} with NFW halos. Although the SIS has a very simple lensing distribution, the 
NFW lens is somewhat more complicated because there are two critical curves 
instead of just one, which leads to infinite magnification at a finite radius in the 
source plane and at the origin (the latter is also true of the SIS).  We have solved 
the general lensing equation in the source plane for a range of lens and source 
redshifts, determining the locations of the images and calculating $\mu$ for each 
one.  Our calculations show that spherical NFW profiles have slightly larger 
magnifications than SIS profiles for an equivalent Einstein angle.  Our SIS lensing 
volumes should therefore be taken to be lower limits.

\begin{figure*}
\begin{center}
\begin{tabular}{cc}
\epsfig{file=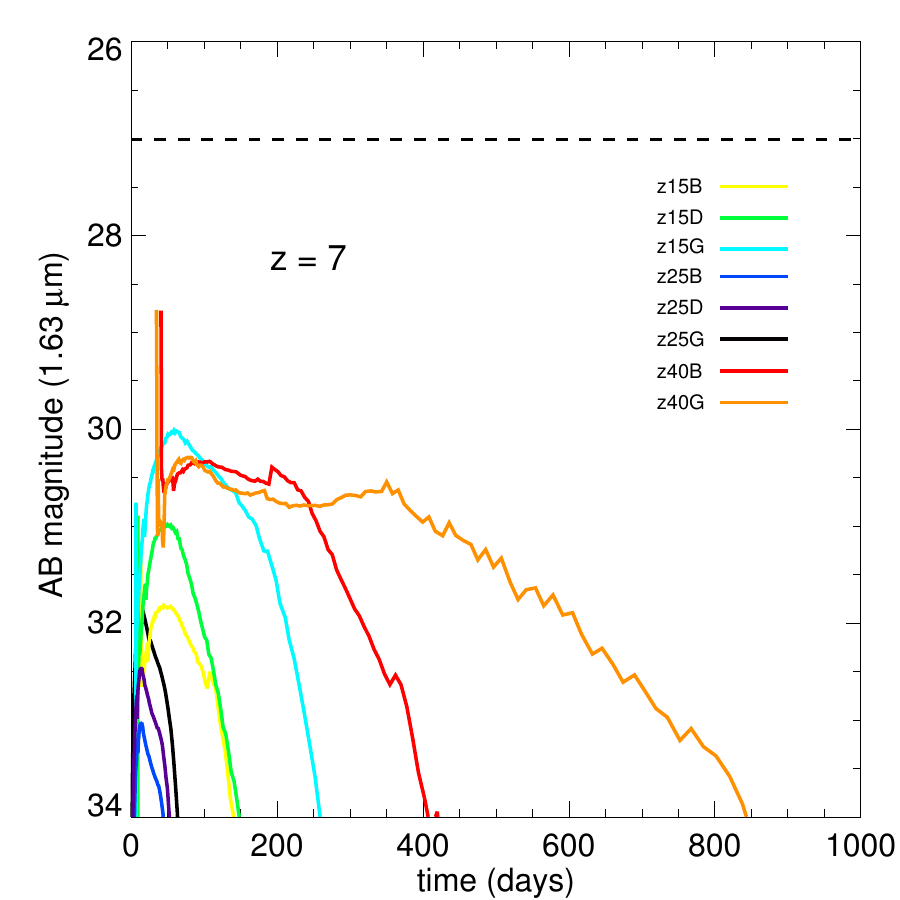,width=0.41\linewidth,clip=} & 
\epsfig{file=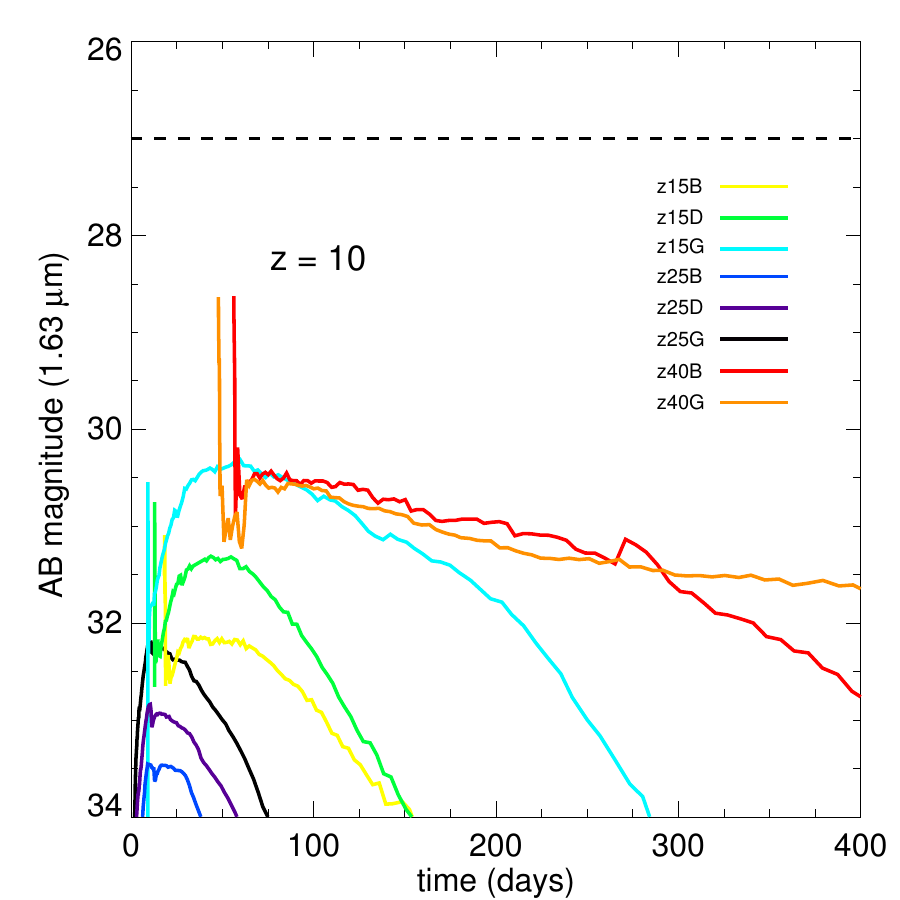,width=0.41\linewidth,clip=} \\
\epsfig{file=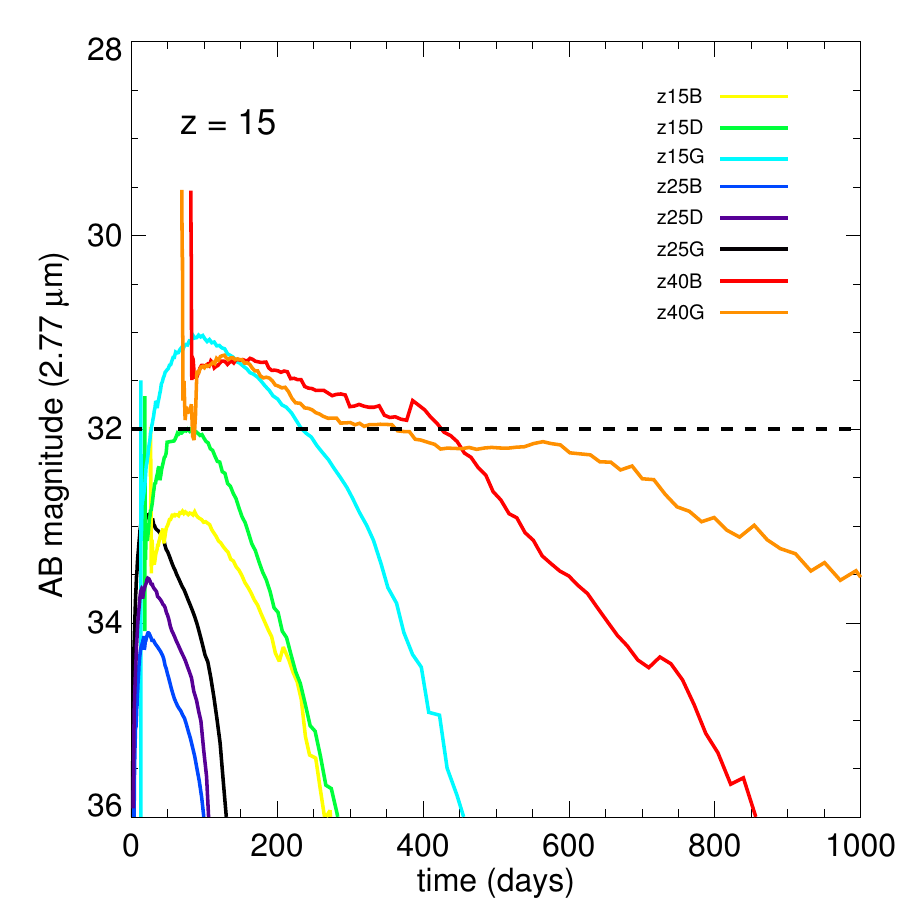,width=0.41\linewidth,clip=} &
\epsfig{file=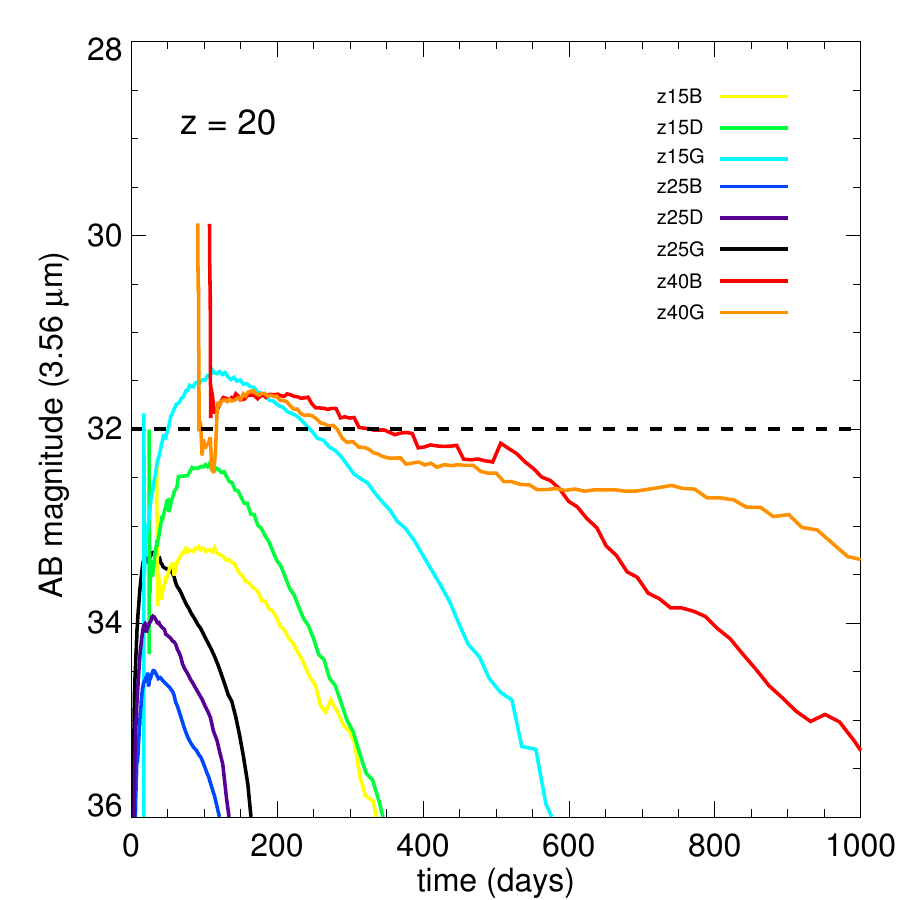,width=0.41\linewidth,clip=}
\end{tabular}
\end{center}
\caption{\textit{CLASH} (1.63 $\mu$m) and \textit{JWST} NIRCam (2.77 and 3.56 
$\mu$m) light curves for 15 - 40 \Ms\ Pop III CC SNe at $z =$ 7, 10, 15, and 20.  
The \textit{JWST} filters shown are those in which the SN is brightest at that 
redshift.  Times on the $x$-axes are in the observer frame.  The dashed horizontal 
lines at AB mag 27 and 32 are \textit{CLASH} and \textit{JWST} detection limits, 
respectively. \vspace{0.1in}}
\label{fig:nir}
\end{figure*}

\section{SN Light Curves}

In Fig.~\ref{fig:nir} we show H band light curves for 15 - 40 \Ms\ Population III 
(Pop III) CC SNe at $z =$ 7 and 10 in the \textit{HST} Wide-Field Camera 3 
(WFC3) 1.63 $\mu$m filter, whose sensitivity limit is $\sim$ AB mag 27.3 for 
total exposure times of a few 10$^5$ s.  These explosions are designated by 
progenitor type (z-series are explosions of red supergiant stars), mass (15, 
25 and 40 \Ms) and energy (B, D, and G are 0.6, 1.2 and 2.4 $\times 10^{51}
$ erg) \citep{wet12c}.  Our SNe and spectra were modeled with the Los 
Alamos RAGE and SPECTRUM codes \citep{fet12}.  The spectra were then 
cosmologically redshifted, dimmed, and convolved with absorption by the 
neutral IGM and {\it HST} and {\it JWST} filter response functions to obtain NIR 
light curves.  

What mostly determines the light curve of a CC explosion is its energy and the 
internal structure of the star.  Rotational and convective mixing lead to similar 
structures for zero- and solar-metallicity stars of equal mass.  Because the 
energy of the SN depends on the entropy profile of the core of the star prior to 
explosion, it also does not vary strongly with metallicity \citep[see Fig.~1 of][]{
wf12}.  Furthermore, at zero metallicity most 15 - 40 \Ms\ stars are thought to 
die as red supergiants because of internal mixing over their lives.  For these 
reasons, our light curves are likely to be representative of CC SNe over all the 
redshifts in our study, not just the earliest epochs.  We consider only CC SNe 
because PI and Type IIn SNe need little or no magnification to be seen by {\it 
CLASH} and they are much less frequent than Type II SNe.
   
Unlike the fiducial 15 \Ms\ CC SN in \citet{pl13}, our SNe have H band fluxes 
that can be detected by {\it CLASH} with only moderate $\mu =$  10 - 20 at 5 
$< z < 12$.  At $z= $ 15 there is essentially no flux at 1.63 $\mu$m so {\it HST} 
would not detect Type II events above this redshift.  At $z =$ 15 they are 
brightest in the 2.77 $\mu$m {\it JWST} NIRCam filter, as shown in 
Fig.~\ref{fig:nir}. At this redshift, three of the SNe exceed the NIRCam detection 
limit of AB mag 32 without being lensed, and $\mu$ of only $\sim$ 2 - 5 are 
required to boost the others above this limit.  Only modest magnifications would 
be required to boost the z15B, z15D, z15G, z40D and z40G explosions above 
the detection limit for {\it JWST} at $z =$ 20, so a cluster lens could in principle 
reveal CC SNe from this epoch.  But small lensing volumes and low star 
formation rates above $z \sim$ 13 make it unlikely that a cluster survey would 
encounter such events, as we discuss below.
 
\section{High-Redshift Star Formation Rates}

\begin{figure}
\epsscale{1.2}
\plotone{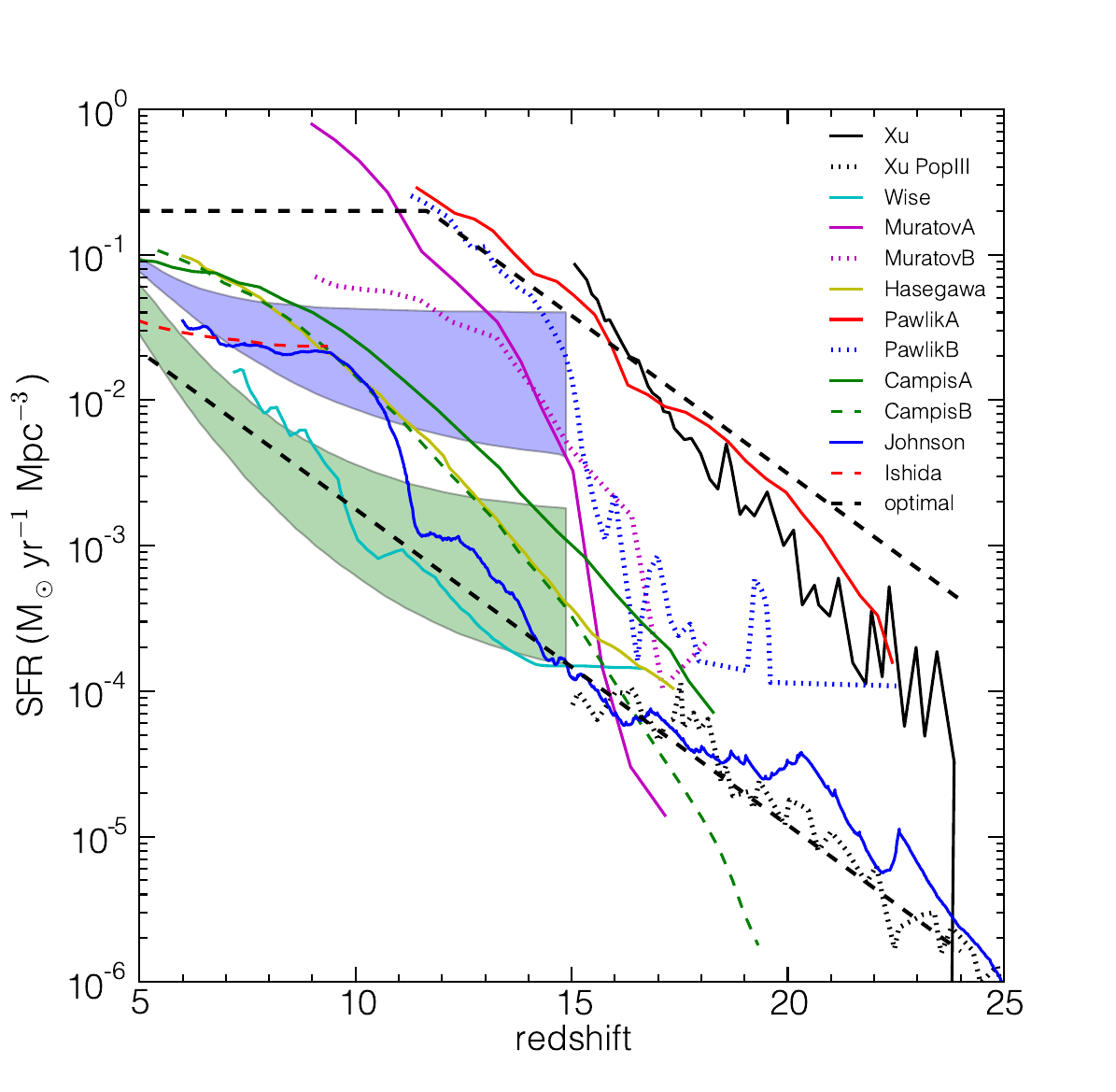}
\caption{SFRs over cosmic time. The blue and green bands are rates inferred from 
GRBs \citep{idf11,re12} and luminosities of high $z$ protogalaxies \citep{camp11},
respectively. The SFRs compiled from numerical simulations are from \citet{wise12,
jdk12,pmb12,haseg13,xu13,mura13}. The two dashed lines are mean and upper fits 
to the simulation SFRs.  The Xu Pop III SFR is for Pop III stars only, the others are 
total Pop III $+$ II SFRs.
\vspace{0.1in}}
\label{fig:sfr}
\end{figure}

We show star formation rates (SFRs) inferred from the luminosity functions of 
early galaxies \citep{camp11}, from GRB rates \citep{idf11,re12}, and from 
numerical simulations \citep{wise12,jdk12,pmb12,xu13,haseg13,mura13} for 
5 $< z <$ 25 in Fig.~\ref{fig:sfr} \citep[see][for other SFRs from simulations that 
fall within the range shown here]{tfs07,ts09}.  The SFRs from early galaxies 
exclude the steep faint end slopes of low luminosity galaxies and are likely 
lower than the true rates.  The SFRs from \citet{re12} and \cite{idf11} above $z 
\sim$ 5 - 7 are mostly extrapolated from those at lower redshifts.  For SFRs 
above $z \sim$ 15 we must rely entirely upon simulations.

SFRs from GRBs vary by $\sim$ 50\% at $z = 5$ and by a factor of 10 at $z =$ 15.  
The rates from early galaxies vary by the similar factors over the same redshifts, 
with an overall spread in SFR between the two methods of a factor of $\sim$ 400 
at $z =$ 15.  The rates predicted by numerical simulations also vary by about two 
orders of magnitude from $z \sim$ 10 - 25.  The SFRs in most of these models are 
just those in a few protogalaxies in their simulation volumes and are therefore 
subject to small box statistics.  The simulations also include a variety of feedback 
processes that lead to a wide range of evolution in stellar populations over cosmic 
time.  We sample this broad range of SFRs with three fits.  On the low end is the 
dashed line in Fig.~\ref{fig:sfr} that borders the simulations with the lowest SFRs 
from below.  As a middle case, we take the dashed line through the simulations 
with the higher SFRs down to $z =$ 15, where it intersects with the GRB rates, 
and then adopt the GRB rates down to $z =$ 5.  As an upper case, we take the 
upper dashed line down to $z \sim$ 12, where the SFR is 0.2 \Ms\ yr$^{-1}$ Mpc$
^{-3}$, and then level off the SFR at this rate down to $z =$ 5.  This rate is equal 
to SFRs inferred from both GRBs and SN measurements at $z \sim$ 3 \citep[for 
the latter, see][]{strolg04}, and is a conservative upper limit.

\section{SN Snapshot Rates}

\begin{figure*}
\plottwo{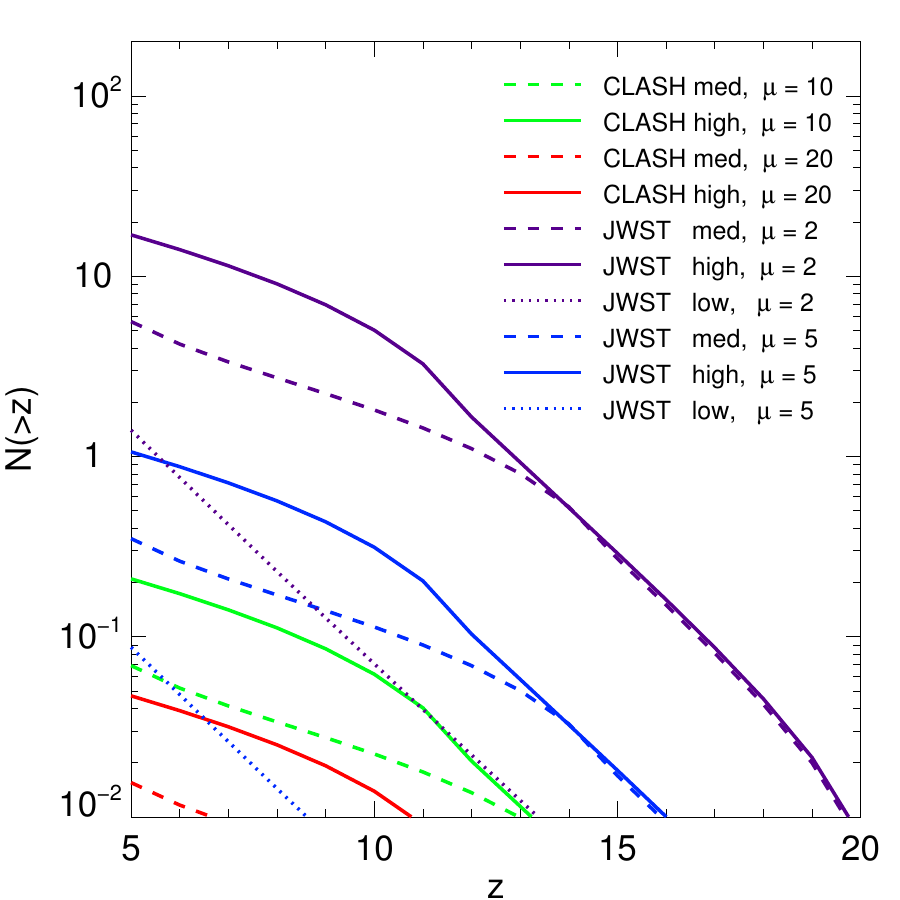}{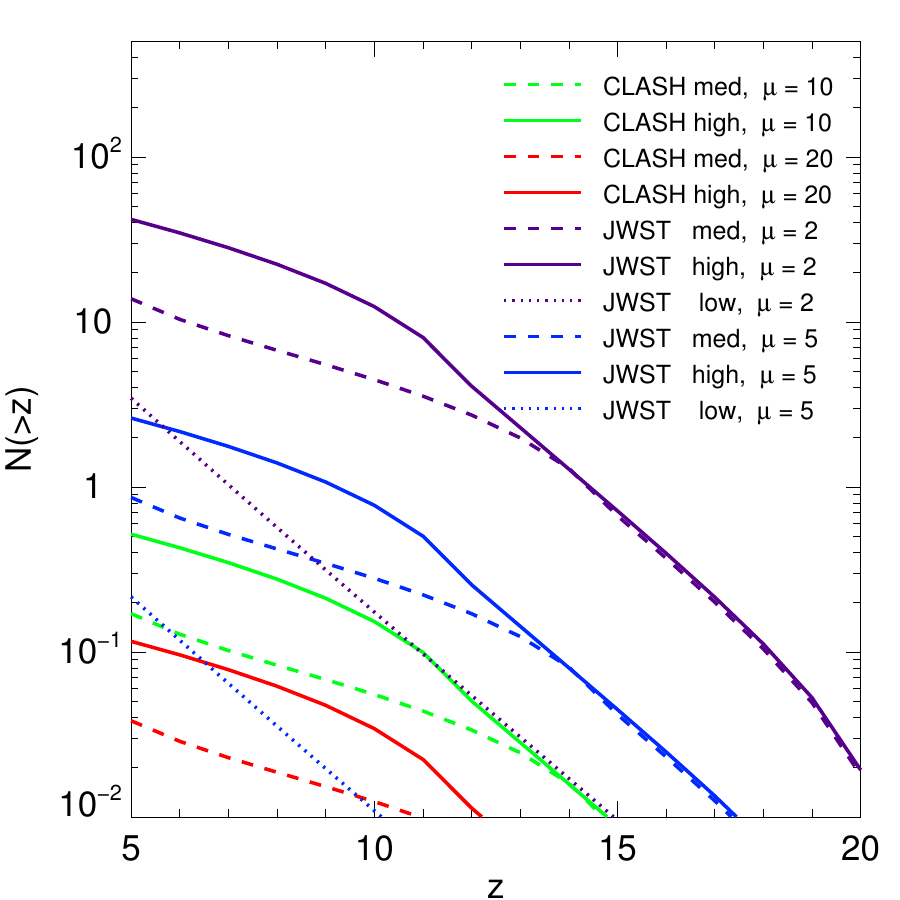}
\caption{The number of SNe detected above a redshift $z$ by a single 35 arcsec 
lens (left) and a 55 arcsec lens (right) in \textit{CLASH} and by \textit{JWST}.  
\vspace{0.1in}}
\label{fig:nsne}
\end{figure*} 

The SN rate can be derived from the cosmic SFR if the initial mass function (IMF) and 
upper and lower mass limits for CC SN production are known.  By $5 < z < 12$, most 
stars form in young galaxies in which metals and dust are produced, so we adopt the 
Salpeter IMF and the usual rule of thumb of $\sim$ 1 SN / 100 \Ms\ of stars formed for 
this mass distribution.  This IMF underestimates the SN rate above $z \sim$ 12 
because a greater fraction of SF likely goes into stars above 20 \Ms, so our SN rates 
should be taken as lower limits.  We have verified that our choice of SN rate per solar 
mass does not vary by more than a factor of two for reasonable choices of upper and 
lower masses for CC SNe and our IMF.  This is far less than the uncertainty in the 
cosmic SFR itself, which is up to a factor of $\sim$ 400. For simplicity, we also take the 
3 or 4 brightest of our SNe to represent CC SNe at high redshift in general.

If SFR is the rate of star formation observed today and d$t$ is the cadence, or time 
between observations of a given cluster, the rate and time between observations at 
redshift $z$ are SFR $(1 + z)$ and d$t / (1 + z)$, respectively. The number of SNe 
enclosed by the lensing volume $\mathrm{d}V(z,\mu)$ whose flux is boosted by a 
factor $\mu$ or more at redshift $z$ over the time d$t$ in the observer frame is then 
\vspace{0.05in}
\begin{equation}  
\mathrm{d}N(z,\mu) \; = \; \mathrm{SFR}  \; (1 + z)  \; \frac{1  \; \mathrm{SN}}{100  \; \Ms}  
\; \mathrm{d}V(z,\mu) \frac{\mathrm{d}t}{1 + z}. \vspace{0.05in}
\label{eq:nsn}
\end{equation}  
To obtain the total number of SNe above a given redshift $N(>z)$ whose flux would be 
boosted above {\it CLASH} or {\it JWST} detection limits, we integrate Eq.~5 over all $z$ 
above that redshift and all $\mu$ above the minimum magnification needed to detect the 
event at each redshift.  We assume an average cadence d$t =$ 3 months and take $\mu
_{min}$ to be $\sim$ 10 for \textit{CLASH} at $5 < z < 12$ for our SNe because their peak 
AB magnitudes do not evolve strongly over this range.  We assume $\mu_{min} \sim$ 2 
for $5 < z < 20$ for \textit{JWST} as a conservative estimate, since even at $z \sim$ 20 
some of the SNe do not have to be lensed to be detected.  The total number of SNe that 
would be found by 35 and 55 arcsec lenses in {\it CLASH} and by {\it JWST} are plotted in 
the left and right panels of Fig.~\ref{fig:nsne}, respectively.  The SFRs used in each plot 
are designated by "hi" (high), "med" (medium), and "low" (low), the three fits described at 
the end of Section 4.  We include curves for $\mu =$ 20 for {\it CLASH} and for $\mu =$ 
5 for {\it JWST} to consider dimmer events.

For (very) low SFRs, the number of SNe that {\it CLASH} would detect in both lenses is 
too small to appear on these plots.  For the medium rates at $\mu =$ 10, the 35 and 55
arcsec lenses would find 0.08 and 0.2 SNe, respectively, or 1 - 3 events total above $z 
=$ 5.  For the high rates, the two lenses would detect 0.2 and 0.5 SNe, or 3 - 9 events 
at $z >$ 5 in the entire survey.  If all the SNe at this epoch were dimmer events that 
require magnifications of 20 to be seen, at most one or two would be detected by {\it 
CLASH}.  The practical limit to CC SN detections with {\it CLASH} is $z \sim$ 12 - 13; 
above these redshifts, the number of SNe that either lens would detect is too low for 
any to appear in the survey, even at high SFRs.  However, at SFRs that are between 
the medium and high rates, one or two SNe in the survey could be $z \sim$ 10 - 12 
events.  The total number of CC SNe that may be found in {\it CLASH} is similar to the 
number of PI SNe that might be found by {\it JWST} at $z \sim$ 20 over its mission life 
\citep{hum12}. 

Future cluster surveys by \textit{JWST} that are similar in scope to \textit{CLASH} will 
detect far greater numbers of CC explosions at 5 $< z <$ 12.  For the medium SFRs at
$\mu =$ 2, the 35 and 55 arcsec lenses would detect 18 and 40 events each.  Unlike
{\it CLASH}, {\it JWST} would detect events even at the lowest SFRs:  at $\mu =$ 2, the 
two lenses could observe 1.5 and 3 SNe each. Even 1 - 2 dimmer events would appear
in such surveys at these rates.  Furthermore, {\it JWST} would find 1 - 3 CC SNe at $z 
\sim$ 15 - 17, almost the era of first star formation.  

\section{Conclusion}

Although primordial SNe won't be discovered prior to the arrival of {\it JWST}, WFIRST 
and the next generation of extremely large telescopes, our calculations show that {\it 
CLASH} may already have found a few SNe (and perhaps up to 10) at 5 $< z < $ 12, 
redshifts well above those at which SNe have been found to date \citep[$z =$ 3.9;][]{
cooke12}.  Such detections could probe the properties of stars in the first galaxies and 
constrain the cosmic SFR at 5 $ < z < 12$ to within the uncertainty of the IMF over this 
interval, which is much smaller.  Even the failure to detect SNe in {\it CLASH} would be 
useful because it would place upper limits on SFRs at 5 $ < z < 12$ by ruling out the 
models with the highest rates.   Our results argue for the use of longer cadences in 
future surveys of cluster lenses, both to capture more events and more easily identify 
explosions with slowly evolving luminosities as transients.  Any doubt over whether a 
particular event is actually a SN could easily be resolved by additional observations of 
that cluster.

Future surveys by {\it JWST} that are similar in scope to {\it CLASH} could discover 
hundreds of SNe out to $z \sim$ 15 - 17, but not at higher redshifts because event 
rates are so low and the lensing volumes with the $\mu$ needed to reveal the SNe 
are so small.  An alternate approach might be to exploit the strong lensing of events 
everywhere on the sky at some redshift by all the structures below that redshift in 
future wide-field surveys.  Far greater volumes at high redshifts could be lensed in 
this manner than by galaxy clusters, albeit at lower average magnifications, and  
compensate for the low SFRs at $z \gtrsim$ 15.  This approach would be ideally 
suited to future all-sky campaigns by WFIRST and WISH.  It is not yet known if the 
lower peak $\mu$ in all-sky lensing, which are due to the smaller masses of lenses 
at high $z$, will be sufficient to boost flux from ancient SNe above the detection 
thresholds of these missions.  Calculations are now underway to determine if strong 
lensing in all-sky surveys will open yet another window on the primordial universe.
  
\acknowledgments

DJW thanks Utte Rydberg and Michele Trenti for helpful discussions in the course of 
work.  He was supported by the Baden-W\"{u}rttemberg-Stiftung by contract research 
via the programme Internationale Spitzenforschung II (grant P- LS-SPII/18).  JS was
supported by a LANL Director's Postdoctoral Fellowship.  Work at LANL was done 
under the auspices of the National Nuclear Security Administration of the U.S. Dept of 
Energy at Los Alamos National Laboratory under Contract No. DE-AC52-06NA25396.  


\end{document}